\documentclass[aps,prb,floatfix,twocolumn]{revtex4}
\usepackage{graphicx}
\newcommand{\BEQ}{\begin{equation}}
\newcommand{\EEQ}{\end{equation}}
\newcommand{\BEA}{\begin{eqnarray}}
\newcommand{\EEA}{\end{eqnarray}}

\renewcommand{\d}{{\rm d}}

\begin{document} 
\title
{Force-induced unzipping of DNA with long-range correlated sequence}
\date{\today}
\author{A.E. Allahverdyan$^{1}$ and Zh.S. Gevorkian$^{1,2}$}
\address{$^{1}$Yerevan Physics Institute,
Alikhanian Brothers St. 2, Yerevan 375036, Armenia.\\
$^{2}$ Institute of Radiophysics and Electronics, Ashtarak-2 378410, Armenia }


\begin{abstract}
  We consider force-induced unzipping transition for a heterogeneous
  DNA model with a long-range correlated base-sequence. It is shown
  that as compared to the uncorrelated situation, long-range
  correlations smear the unzipping phase-transition, change its
  universality class and lead to non-self-averaging: the averaged
  behavior strongly differs from the typical ones.  Several basic
  scenarios for this typical behavior are revealed and explained.
  The results can be relevant for explaining the biological purpose
  of long-range correlations in DNA.

\end{abstract}
\pacs{
PACS: }
\maketitle


{\it Introduction.} Structural transformations of DNA under changing
of external conditions are of primary importance for molecular biology
and biophysics \cite{bio}. They take place in transcription of genetic
information from DNA and in duplication of DNA during cell division
\cite{bio}.  The common physical scenario of both these processes is
unwinding of the double-stranded structure of DNA under influence of
external forces.  Whereas theoretical and experimental studies of
thermal denaturation (melting) of DNA have a long history,
force-induced unzipping has been actively investigated only relatively
recently \cite{hindu,nelson,cocco,exp}; for a concise review and more references
see \cite{cocco}. The research in this field is motivated by the new
generation of micromanipulation experiments \cite{exp}. For the
theoretical understanding of the subject both homo and heteropolymer
models of DNA were studied \cite{hindu,nelson,cocco}.

The main purpose of the present letter is to make the next step
towards real DNAs and to analyze force-induced unzipping for a
DNA-model, where the correlation structure of the base-sequece is
taken into account. Indeed, one of the main differences between DNA
and other biopolymers \cite{jenya} is that the base-sequence of the
former displays long-range correlations (1/f-noise spectrum)
\cite{longrange,voss,chin}; for a review see \cite{review}.  Recall
that the base-sequence of a DNA molecule consists of purines (A and G)
and pyrimidines (C and T).  They constitute the genetic code carried
by DNA. Initial studies reported long-range correlations for
non-coding regions of DNA, while more recent results show that certain
types of them can also exist in coding regions \cite{coding}.
Moreover, systematic changes were found in the structure of
correlations depending on the evolutionary category of the DNA carrier
\cite{voss,chin}. In spite of ubiquity of long-range correlations in
DNA-structures, their biological reason remains basically unexplored.

We will show below that long-range correlations present in the
base-sequence of DNA make its behavior under the unzipping external
force essentially non-self-averaging: there are several widely
different scenarios of behavior which specifically depend on the
concrete structure of the base-sequence and are not reproduced by
the averaged behavior.  This is in contrast to DNAs with short-range
correlated base-sequence whose behavior in the vicinity of the
unzipping transition is perfectly self-averaging: almost every
molecule behaves (in the thermodynamic limit) similar to the average.

{\it The model} we will work with takes into account the most minimal
amount of physical ingredients needed to describe force-induced
unzipping. {\it i)} a DNA molecule is lying along the $x$-axis between
the points $x=a$ and $x=L$. {\it ii)} only inter-strand (hydrogen)
bonds of the molecule are considered; they are located at points
$x_i$, $a<x_i<L$, $i=1,...,M$. Any bond can be in one of two states:
bound or broken. We choose the overall energy scale in such a way that
the latter case contributes to the Hamiltonian a binding energy
$\phi(x_i)$, whereas the former case brings nothing.  Different types
of bonds do have different binding energies, so $\phi(x_i)$ is a
random quantity with an average $\langle\phi\rangle$:
$\phi(x_i)=\langle\phi\rangle+\eta(x_i)$. {\it iii)} a force is acting
on the left end $x=a$ of the molecule pulling apart the two strands.
Thus, if a bond $x_i$ is broken, all the bonds $x_j$ with $j<i$ are
broken as well. Each broken bond brings additionally to the
Hamiltonian a term $-{\cal F}$, where ${\cal F}$ is proportional to
the acting force. {\it iv)} summarizing all of these, one comes to the
Hamiltonian $H(x)=-{\cal F}x+\sum_{i=1}^x
\phi(x_i)=(\langle\phi\rangle-{\cal F})x+\sum_{i=1}^x \eta(x_i)$,
where $x$ is the number of broken bonds. In the thermodynamical limit,
where $L$ and $M$ are large, one applies the continiuum description
with $x$ being a real number, $a<x<L$, and ends up with the following
Hamiltonian and partition function:
\BEA
\label{d1}
H(x)=f(x-a)+\int_a^x\d s\,\eta(s),~
Z=\int_a^L\d x\, e^{-\beta H(x)},
\EEA
where $f=\bar{\phi}-{\cal F}$ and $\beta=1/T$ is the inverse temperature
($k_{\rm B}=1$). It remains to specify the properties of the noise
$\eta$. Strictly speaking, it can take values corresponding
to inter-strand bonds AT and GC. However, within the adopted
description we assume it is a gaussian stationary process with an
autocorelation function $K(t-t')=\langle\eta(t)\eta(t')\rangle$ 
to be specified later on. The model given by
(\ref{d1}) and by $K(t)\propto\delta(t)$ (white noise) is
well-known, and was used to describe interfaces, random walks in a
disordered media, and aspects of population dynamics
\cite{manfred}. It was recently applied for the unzipping transition
in DNA \cite{nelson}.

{\it Reduction to a stochastic differential equation.} In
Eq.~(\ref{d1}) one fixes $L$, and views $a$ as a parameter varying
from the highest possible value $L$, where $Z=0$, to the lowest
possible value which we define to be $a=0$. The quantity $t=-a$ will
thus monotonicaly increase and can be interpreted as a time-variable.
Differentiating $Z$ in (\ref{d1}) over $a$ and changing the variable as
$t=-a$, one gets:
\BEA
\label{d2}
\frac{\d Z}{\d t}=1-\beta f Z-\beta\,\eta(t)Z,
\qquad -L<t<0
\EEA
where we used $\eta(t)=\eta(-t)$, as follows from the gaussian stationary
property of the noise. This is a Langevin equation with
a multiplicative noise. From (\ref{d2}) one can obtain a stochastic equation for
$F=-T\ln Z$:
\BEA
\label{d3}
\frac{\d F}{\d t}+V'(F)=\eta(t),\quad
V(F)=T^2e^{\beta F}-fF.
\EEA
The {\it order parameter} of the problem is the number of
broken bonds. Along with its average and variance it is defined 
for $t=0$ as 
\BEA
\label{brams}
x=\partial_fF,\quad
\overline{x}=\partial_f\langle F\rangle,\quad \overline{\Delta x^2}\equiv
\overline{x^2}-\overline{x}\,^2=-T\partial_f \overline{x}.
\EEA

{\it Exponentially correlated noise.} As the first example we shall
consider Ornstein-Uhlenbeck (OU) noise 
$K(t)=(D/\tau)\,e^{-|t|/\tau}$, 
where $D$ is the intensivity, and $\tau$ is the correlation
time. Although for a finite $\tau$ this noise is short-range
correlated, we believe that it correctly catches the basic trends of
the more general situation when changing $\tau$ from $0$ to some large
value. Note that the white-noise situation is recovered for $\tau\to 0$.

To handle (\ref{d3}) one differentiates it over $t$ and uses the
generating equation for OU process
$\tau\dot{\eta}=-\eta+\sqrt{D}\xi(t)$, where $\xi(t)$ is a white
gaussian noise: $\langle\xi(t)\xi(t')\rangle=2\delta(t-t')$.
Introducing $s=t/\sqrt{\tau}$ one gets \cite{hanggi}:
\BEA
\label{bukhara}
\frac{\d ^2 F}{\d s^2}+\gamma(F)\,
\frac{\d F}{\d s}=-V'(F)+\frac{\sqrt{D}}{\tau^{1/4}}\xi(s),
\EEA
where $\gamma(F)=1/\sqrt{\tau}+\sqrt{\tau}\,V''(F)$.  Recall that
Eq.~(\ref{bukhara}) has the same form as a Langevin equation for a
particle with unit mass in the potential $V(F)$ and subjected to a
white noise and a $F$-dependent friction with a coeffcient
$\gamma(F)$. $V(F)$ is confining only for $f>0$: $V(F)\to \infty$
for $F\to\pm\infty$. As well-known
\cite{risken}, for sufficiently long times one can neglect the
inertial term $\d ^2 F/\d s^2$, provided that at least one of the
following conditions are satisfied: {\it (i)} the dependence on $F$ in
$\gamma(F)$ is weak; {\it (ii)} $\gamma(F)$ is sufficiently large.  If
$V''(F)$ is of order one, then the second condition is satisfied both
for large and small $\tau$ \cite{hanggi}. If $V''(F)$ is small then
the first condition is satisfied.  After neglection of the inertial
term in (\ref{bukhara}), the remainder is an ordinary white-noise
Langevin equation, and, by means of standard methods \cite{risken},
can be transferred to a Fokker-Planck equation for the distribution
function $P(F,s)=\langle\delta(F-F[s])\rangle$, where $F[s]$ is a
particular, noise-dependent solution of (\ref{bukhara}).
\BEA
\frac{\partial P}{\partial s}-\frac{\partial }{\partial F}
\frac{V'(F)}{\gamma(F)}P(F)=\frac{D}{\sqrt{\tau}}\,
\frac{\partial }{\partial F}
\frac{1}{\gamma(F)}
\frac{\partial }{\partial F}
\frac{P(F)}{\gamma(F)}.
\label{samarkand}
\EEA
For large times (lengths), i.e. for $L\gg 1$ 
and $t\propto s\to 0$, any solution of (\ref{samarkand}) tends to
the stationary distribution (see e.g. \cite{risken} for a general
proof) obtained from (\ref{samarkand}) by putting $\partial _sP=0$:
\BEA
 P_{\rm st}(F)={\cal N}\,\gamma(F)\exp\left[-\frac{\tau}{2D}[V'(F)]^2
-\frac{1}{D}V(F)\right],
\label{trick}
\EEA
where ${\cal N}$ is the normalization factor. 
The white-noise, $\tau\to 0$, limit of $P_{\rm st}(F)$ was obtained in
\cite{manfred,nelson}. 

The critical domain of the model coresponds to $f\to +0$, where the
average energy cost for breaking a hydrogen bond tends to zero. Our
aim is to compare in this domain the behavior of $\overline{x}$ for a
finite $\tau$ with that of $\tau=0$ (white-noise) as to determine the
effect of the noise-correlation. Recall from \cite{manfred} that for
$\tau=0$ a simple formula exists: $\overline{x}=T^2\psi'(\mu)$, where
$\mu=Tf/D$ and $\psi'(\mu)=\d ^2[\ln\Gamma(\mu)]/\d\mu^2$. Thus for
$f\propto\mu\to 0$, $\overline{x}$ becomes large \cite{nelson}:
$\overline{x}=D/f^2$. One can explain this by noting that for $f\to 0$
the potential $V(F)$ in (\ref{d3}) ceases to be confining, and the
particle escapes to infinity. Note that here the random quantity $x$
is concentrated around its average: $\overline{\Delta x^2}/
\overline{x}\,^2\propto f\to 0$. Thus, in the present context a given
DNA molecule with a typical base-sequence does not have 
individuality: its behavior coincides with the averaged one.

\begin{figure}[bhb]
\includegraphics[width=7cm]{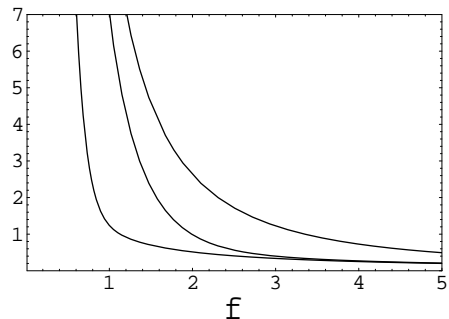}
\vspace{-0.5cm}
\caption{ The order parameter $\overline{x}$ versus $f$
for Ornstein-Uhlenbeck noise with $D=10$, $T=1$. 
From right to left: $\tau=0$, 
$\tau=10$, $\tau=100$. }
\label{f1}
\end{figure}

$P_{\rm st}(F)$ and $\overline{x}$ can be
expressed via Kummer functions and then easily 
studied numerically.  Fig.~(\ref{f1}) shows that although
the behavior of $\overline{x}$ for very small $f$ does not depend much
on $\tau$, such a dependence does exist for moderately small values of
$f$: finite $\tau$'s smear the small-$f$ singularity and thus increase
the stability of the DNA molecule, since larger external forces ${\cal
F}$ are demanded to achieve the same amount of broken bonds.

{\it Long-range correlations} will be modelled via a 
stationary gaussian noise with an autocorrelation function \cite{chin}:
\BEA
\label{korund}
K(t)\equiv\langle\eta(t)\eta(0)\rangle =\sigma |t|^{-\alpha},
\EEA 
where $0<\alpha<1$ is the exponent characterizing the long-range
correlation, and where $\sigma$ is the intensivity.
Although the real noise distributions in DNA can be much more
complicated \cite{review,coding}, 
(\ref{korund}) is certainly the minimal model of noise which 
allows to study long-range correlations.

To start with, lest us consider a case with $\alpha=0$,
which does not have a direct physical interest and does not allow the
strict thermodynamical limit, but, as seen later, it is still able to
provide a relevant insight. The noise is now completely frozen:
$\eta(s)$ in (\ref{d1}) does not depend on $s$, and due to this the
problem is easily solved from (\ref{d1}):
\BEA
\label{karma}
\frac{\bar{x}}{L}=\left\langle g(\eta)\right\rangle,\quad
g(\eta)\equiv \frac{T}{L(f+\eta)}-\frac{1}{e^{\beta L(f+\eta)}-1},
\EEA where $\langle ...\rangle$ is taken over the
zero-average gaussian random quantity $\eta$ whose dispersion is
$\sigma$.  If $\beta L$ is large, $g(\eta)$ behaves as the
step-function, $g(\eta)\simeq \theta(-\eta-f)$: for any single
realization of the noise there is a sharp phase
transition with a jump at the realization-dependent point $f=-\eta$
(non-self-averaging).  In contrast, due to the integration over $\eta$
in (\ref{karma}), the behavior of $\overline{x}$ is smooth, and there
remains only a crossover between small $\bar{x}/L$ for a large $f$ and
$\bar{x}=L/2$ for $f=0$: the sharp transition disappears.

We return to Eq.~(\ref{d2}) with the noise $\eta(t)$ characterized by
(\ref{korund}). Our aim is to obtain a Fokker-Planck equation for the
probability density $P(Z,t)=\langle\,\delta(Z-Z[t])\,\rangle$,
where $Z[t]$ is a noise-dependent solution of (\ref{d2}).
Differentiating $P(Z,t)$ over $t$ and using (\ref{d2}), one gets:
\BEA
&&\frac{\partial P}{\partial t}+\frac{\partial}{\partial Z}
[\,(1-\beta fZ)P\,]=\beta\frac{\partial}{\partial Z}Z
\langle\eta(t)\delta(Z-Z[t]\,)\rangle.\nonumber\\
&&
\label{krunk}
\EEA
To handle the last term, one uses the fact that the noise is gaussian
and applies Novikov's theorem \cite{fox}, to obtain
\BEA
&&\langle\eta(t)\delta(Z-Z[t]\,)\,\rangle
=\int_{-L}^t\d s\, K(t-s)
\langle\frac{\delta}{\delta\eta(s)}
\delta(Z-Z[t])\rangle
\nonumber\\&&
=-\frac{\partial}{\partial Z}
\int_{-L}^t\d s\,K(t-s)\left\langle\,
\delta(\,Z-Z[t]\,)\,\frac{\delta
Z[t]}{\delta\eta(s)}\,\right\rangle,
\label{novik}
\EEA
where $\delta \,/\delta\eta(s)$ is the variational 
derivative, the equation for which is obtained from (\ref{d2}).
Solving this equation and using
\BEA
Z[s]=\exp\left[\int_t^s\d u\left(\frac{1}{Z[u]}-\beta f
-\beta\eta(u)\right)\right]Z[t],
\label{abu}
\EEA
also obtained from (\ref{d2}), one finally gets
\BEA
&&\langle\,\eta(t)\,\delta(\,Z-Z[t]\,)\,\rangle
=\frac{\partial}{\partial Z}\,Z\,
\int_{-L}^t\d s\,K(t-s)\times\nonumber\\
&&\left\langle\,
\delta(\,Z-Z[t]\,)
\exp\left[-\int_s^t\frac{\d u}{Z[u]}\,\right]\,
\right\rangle.
\label{aragil}
\EEA
Eqs.~(\ref{krunk}, \ref{aragil}) are exact, but since now
approximations have to be applied to get a closed equation for
$P(Z,t)$. Note from (\ref{d2}, \ref{d3}) the following relation valid
in the stationary state: $\langle 1/Z\rangle =\beta f$. For $f\to +0$
this relation can be satisfied only if the corresponding stationary
distribution tends to become non-normalizable due to its large-$Z$
behavior. Thus, we can search for this distribution assuming that the
characteristic values of $Z$ are large. For this one
takes the thermodynamical limit $L\gg 1$, $t\to 0$, makes partial
integration in the RHS of (\ref{aragil}),
uses (\ref{d2}), and gets for $\phi(Z,s)\equiv\int_0^s\,\d
u/Z[\eta,u]$ ($\beta f\ll 1$):
\BEA
\phi(Z,s)=
\frac{s}{Z[\eta,s]}
-\beta\int_0^s\frac{\d u\,u\,\eta(u)}{Z[\eta,u]}
+\int_0^s \frac{\d u\, u}{Z^2[\eta,u]}.
\label{varaz}
\EEA
Now the last term can be neglected due to
the above large-$Z$ property. Assuming additionally that the magnitude
of the noise $\eta(t)$ is small, one can estimate the second term in
the RHS as being at least of order ${\cal O}(1/Z^2)$ and neglect it as well.
For the first term in the RHS of (\ref{varaz}) one uses (\ref{abu})
to express $Z(s)$ by $Z(0)$ which due to the delta-function in 
(\ref{aragil}) can be substituted by $Z$. The noise in the resulting
equation for $\phi$ is again neglected, and then $\phi$ is determined from:
\BEA
\label{jemen}
Z\,\phi(Z,s)=se^{-\phi(Z,s)+\beta fs}.
\EEA
Thus the stationary distribution reads
\BEA
P_{\rm st}(Z)=\frac{{\cal N}}{Z\,{\cal D}(Z)}\,
\exp{\left[
\int^Z\frac{\d u}{u\,{\cal D}(u)}\,\left(
\frac{T}{u}-f\right)
\right]},
\label{port-said}
\EEA
where ${\cal N}$ is the normalization
(the lower limit of integration is not specified, since it can be
absorbed to ${\cal N}$), and where
\BEA
{\cal D}(Z)=\int_0^\infty\d s\,K(s)\,e^{\phi(Z,-s)}.
\label{dz}
\EEA
To study the critical behavior of $\overline{x}$, one needs two
asymptotic regimes found from (\ref{jemen}, \ref{dz}): 
${\cal D}(Z)\propto Z^{1-\alpha}e^{\beta
f(1-\alpha)Z}$ for $\beta f Z\ll 1$, while ${\cal D}(Z)$ 
is constant for $\beta f Z\gg 1$.  Substituting these into
(\ref{port-said}) and selecting the most divergent terms, one gets:
\BEA
\label{shun}
\overline{x}\simeq \frac{\Gamma(\alpha)}{\tilde{\sigma}(\beta
  f)^{\alpha}}\,\ln\frac{1}{\beta f},\quad \kappa\equiv
\frac{\overline{\Delta x^2}}{\overline{x}\,^2}
\simeq\frac{\tilde{\sigma}\,(\beta
  f)^{\alpha-1}}{\Gamma(\alpha)\ln\frac{1}{\beta f}}, \EEA where
$\tilde{\sigma}=\sigma^{(1-\alpha)/(2-\alpha)}$.  Due to the above
weak-noise assumption, (\ref{shun}) represents the leading term of the
small-$\tilde{\sigma}$ expansion.  It is seen that in contrast to the
white-noise situation the behavior of $\overline{x}$ is smeared, and
that $x$ is strongly non-self-averaging quantity: $\kappa\gg 1$ for
$\beta f\ll 1$ (recall \cite{nelson,manfred} that in the white-noise
case: $\overline{x}\simeq f^{-2}$ and thus $\kappa\simeq f^{-1}\to 0$
for $f\to 0$).  Both these results are contrasting to qualitative
predictions made in \cite{nelson}: $\overline{x}\simeq f^{-2/\alpha}$,
$\kappa\simeq f^{-1+2/\alpha}\to 0$, which means that the small-$f$
singularity is stronger and $x$ is even more self-averaging than in
the white-noise case. We think that this discrepancy is due to
inapplicability of the reasonings made in \cite{nelson} for finite
temperatures.

\begin{figure}[bhb]
\includegraphics[width=7cm]{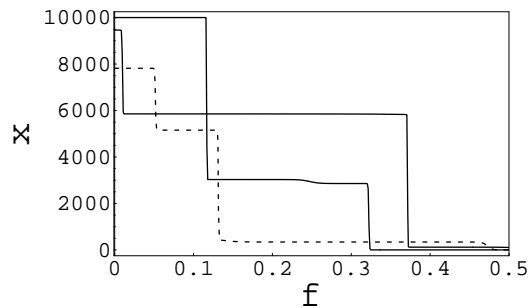}
\vspace{-0.5cm}
\caption{
$x(f)$ for three realizations of the noise 
within another class of typicality.
$T=\sigma=1$, $L=10^{4}$ and $\alpha=0.5$.}
\label{f3}
\end{figure}

{\it Typical scenarios of unzipping.} The above results on
non-self-averaging indicate that 
$\overline{x}(f)$ is not directly relevant for experiments which
are carried out on single DNA molecules: one should study different
realizations of the noise and identify typical, i.e. frequently met,
scenarios of behavior. Results of extensive numerical investigation of
this problem will be reported elsewhere \cite{a}. Here we discuss 
some representative examples.
\begin{figure}[bhb]
\includegraphics[width=7cm]{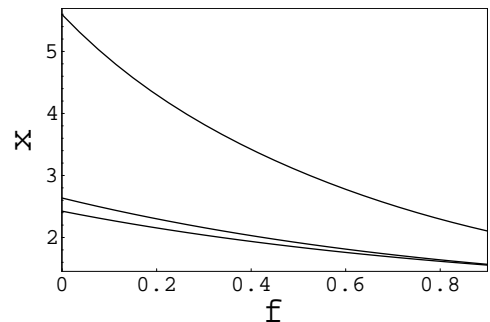}
\vspace{-0.5cm}
\caption{
$x(f)$ for three realizations of the noise
within one class of typicality.
$T=\sigma=1$, $L=10^{4}$ and $\alpha=0.5$.}
\label{f2}
\end{figure}
By means of direct numerical enumeration of $L=10^4$ discrete
base-pairs we studied the behavior of the (unaveraged) order parameter
$x$ as a function of $f$. The long-range correlated gaussian
discrete-time stochastic process was generated following to optimized
recipes proposed in \cite{sancho}. As compared to (\ref{korund}), the
noise was regularized at short distances due to obvious numerical
reasons.  We focus on the thermodynamical domain where $f$ is not very
small, and thus comparison with the theory is possible.  In the
(regularized) white noise case the simulations are in perfect
agreement with the theory: $x$ is self-averaged and
$\overline{x}\propto f^{-2}$ is reproduced. In contrast to that a strong
non-self-averaging is present for the long-range correlated noise.
Moreover, we found several radically different scenarios of the
typical behavior. Two extremal ones among them are presented in
Figs.~\ref{f2}, \ref{f3}. The first one is present in nearly 12\% of all
realizations and is demonstrated by Fig.~\ref{f2}. It is characterized
by very smooth, non-critical behavior of $x(f)$
for $f\ge 0$. Fig.~\ref{f3} presents a strictly different
situation: $x(f)$ increase by several jumps followed by very flat
regions. $x(0)$ is either equal to its maximal possible value $L$ or close
to it. This phase-transition scenario is met in nearly 45\% of all
realizations. Other typical realizations are intermediate between
these two extremes.
Our discussion of the frozen noise made after (\ref{karma}) allows to
explain this jump-plateau structure.  A sizeable portion of long-range
correlated noise realizations can be qualitatively visualized as
several pieces of the frozen noise with different $\eta$ put next to each other.
Now recall from (\ref{karma}) that every
sufficiently long piece of that type has a single first order phase
transition with a jump proportional to its length.

{\it In conclusion}, we have shown that long-range correlations in the
base-sequence of a model DNA drastically influence its unzipping under
external force: {\it i)} the behavior of the average order parameter
in the critical regime is smeared; {\it ii)} the situation is
essentially non-self-averaging: there are several scenarios of typical
unzipping which do not coincide with the averaged behavior; {\it iii)}
long-range correlations increase the adaptability of the molecule,
since in some typical scenarios it becomes more stable with respect to the
force, while in others the unzipping phase transition is amplified.
What scenario will be selected depends on the detailed structure of
the base-sequence.  Some of the above tendencies, e.g. the smearing,
are seen already for a short-range correlated base-sequence. We
hope that these results will contribute into understanding of the
role and the purpose of long-range correlations in DNA.

\end{document}